# Sensitivity of sunspot area to the tidal effect of planet Mercury during solar cycle 23.

**I. R. Edmonds**
12 Lentara St, Kenmore, Brisbane, Australia 4069 ian@solartran.com.au

**Abstract.**
We present evidence that the allowed periods of equatorially trapped Rossby wave modes on the Sun coincide closely with the 88 day period and 176 day sub harmonic period of Mercury and evidence of Rossby waves on the Sun at the same periods. To test the hypothesis that Rossby waves trigger the emergence of sunspots we use band pass filtering to obtain the ~ 88 day period and ~ 176 day period components of hemispherical sunspot area and compare the variations to the tidal height variation on the surface of the Sun due to Mercury. We find that the two components of hemispherical sunspot area occur in several episodes or activations of duration 2 – 6 years during each solar cycle. When the activations are discrete the variation of the ~ 88 day and ~176 day period components are phase coherent with the tidal height variation and a π phase change is evident between successive activations. We use this result to demonstrate that Rieger type "quasi-periodicities" in sunspot activity are, in most reported cases, periodicities associated with sidebands of the 88 day and 176 day period variation. We find that, in variables that arise from activity in both solar hemispheres, phase coherence with the Mercury tidal effect occurs at infrequent intervals but, when it does occur, it is possible to observe phase coherence of variables such as sunspot number, solar flares, solar wind speed and cosmic ray flux with the Mercury tidal effect. Sensitivity of a range of variables to the tidal effect is calculated for such intervals in solar cycle 23.

## 1. Introduction

Periodicities in the range 50 to 200 days in solar activity, in interplanetary space and in terrestrial variables are known as Rieger-type quasi-periodicities, Reiger (1984). The quasi-periodicities are characterised by intermittent periodicity in episodes of 1 – 3 years during epochs of maximum solar activity with periods around 160 days quite common, Richardson and Cane (2005), Ballester et al (2000, 2004). These enigmatic periodicities have been observed in nearly all solar activity and interplanetary space variables. In particular in flares, Reiger et al (1984), Dennis (1985), Ichimoto et al (1985), Landscheidt (1986), Bai & Sturrock (1987), Kyle & Cliver (1991), Bai (1994), Bai (2003), Dimitropolou et al (2008), Bogart & Bai (1985), Kile & Cliver (1991); in proton events, Gabriel et al (1990); in photospheric magnetic flux, Ballester et al (2002), Knaack et al (2005); in radio bursts Verma (1991); in sunspot groups, Lean & Brueckner (1989), Lean (1990), Pap et al (1990), Carbonell & Ballester (1990), Bouwer (1992,) Carbonell & Ballester (1992), Verma and Joshi (1987), Verma et al (1992), Oliver and Ballester (1995), Oliver et al (1998), Ballester et al (1999), and Krivova & Solanki (2002); in interplanetary field, Richardson & Cane (2004); and in cosmic ray flux Mavromichalaki et al (2003), Dragic et al (2008) and Hill et al (2001). While there are a wide range of periodicities reported the most common reference is to the "~150 day quasi-periodicity", Richardson & Cane (2004). Most recently Zaqarashvili et al (2012) have reported time/period diagrams showing the quasi-periodicities in the sunspot area for cycles 19 to 23 indicating the fairly broad range of periods associated with these quasi-periodicities and indicating the strong episodes of periodicity occurring at the maximum of solar cycle 19 (1956 – 1959) and solar cycle 21 (1978 – 1982) in the period range 150 to 170 days.

Early explanations of the quasi-periodicities included Bai et al (1987b) suggesting that the cause must be a mechanism that induces active regions to emit flares, Ichimoto et al (1985) suggesting that the periodicities are associated with the storage and escape of magnetic flux from the Sun, Bai & Cliver (1990) suggesting the periodicities could be simulated with a non-linear damped oscillator, Wolff (1992) suggesting an interaction of modes of oscillation of the Sun, Sturrock & Bai (1993) suggesting that the Sun contains a "clock" with a period of 25.5 days and the periodicities are sub-harmonics of the "clock" period, Wolff (1983) suggesting the interaction of the solar activity band with solar g modes, and Sturrock (1996) suggesting the Sun contains two "rotators" one at ~22 days and the other at ~25 days, that combine to produce the observed periodicities.

More recently Lou (2000) and Lou et al (2003) suggested the periodicities can be linked to equatorially trapped magnetic Rossby waves near the surface of the Sun. Rossby waves or r-mode waves can form in a fluid layer on the surface of a sphere. Characterised by the Laplace tidal equation, Rossby waves also form as large scale waves in the ocean and atmosphere on Earth. Lou (2000) and earlier Wolff (1998) derived dispersion relations for equatorially trapped Rossby waves in the photosphere of the Sun. The amplitude of equatorially trapped Rossby waves vary longitudinally around the equator and have a Gaussian envelope across and centred on the equator. Lou (2000) derived the following expression for the allowed periods, $T(n,m)$:

$T(n,m) = 25.1[m/2 + 0.17(2n + 1)/m]$ days  (1)

where 25.1 days is the sidereal period of rotation of the Sun at the equator, $n = 1, 2$ and $m = n, n+1, n+2 \ldots$ The periods for the $n = 1$ mode i.e. the mode with one nodal line through the poles, are tabulated in Dimitropoulou et al (2008) and are given here, in Figure 1, for both the $n = 1$ and the $n = 2$ modes.

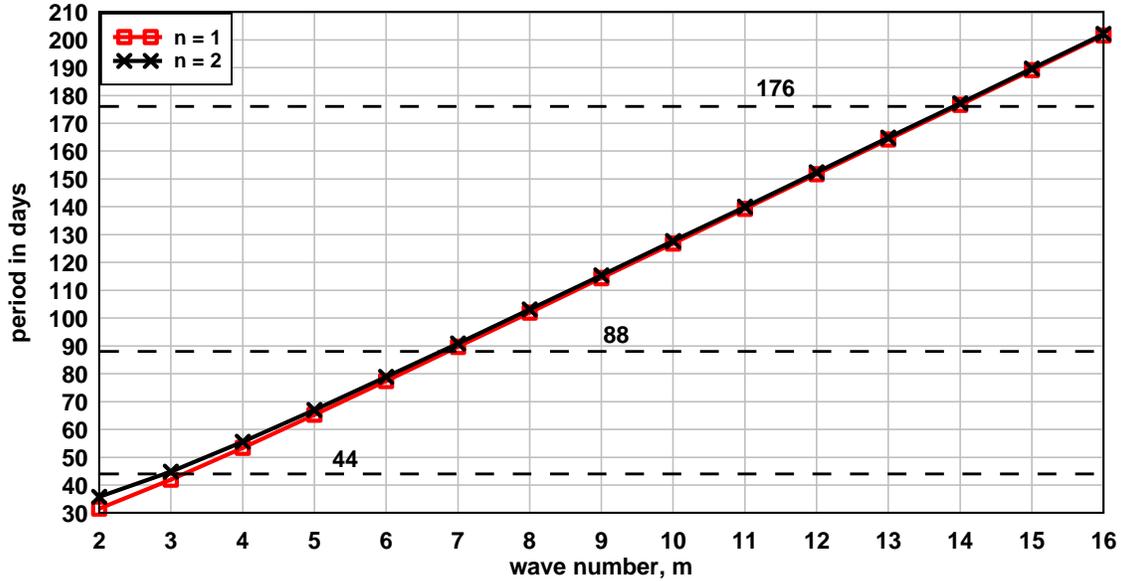

**Figure 1.** The magnetic Rossby wave period from equation (1) for nodal line number n =1 (squares) and n = 2 (crosses) as a function of wave number m. Horizontal lines indicates the period and first harmonic and sub harmonic period of Mercury.

Clearly there is a very small difference between the periods of the 1,m modes and the 2,m modes and the mode periods can be characterised fairly accurately by m alone. Also, as pointed out by Dimitropoulou et al (2008) and evident in Figure 1, the possible modes are quite dense, with a spacing between allowed periods of about 12 days, so that almost all observed Reiger type quasi-periodicities could be attributed to Rossby waves on the Sun. Ballester et al (2002) questioned why the quasi-periodicity in solar activity around 160 days is frequently observed whereas the other possible wave mode periods available from equation (1) are less manifest in solar activity. Dimitropoulou et al (2008), in their concluding remark, noted that "an additional mechanism should be considered, which practically promotes specific modes out of the theoretically equivalent ones." See also the concluding remarks of Ballester et al (2002).

We consider the possibility that the "additional mechanism that promotes specific modes" is the tidal effect of the planet Mercury. In this paper the tidal effect is taken to be the height difference between high and low tide on the surface of the Sun. The height difference or tidal elongation due to planet of mass M and orbital radius R at the surface of the Sun is $(3/2)(r_{Sun}^4/M_{Sun})(M/R^3)$ where $r_{Sun}$ is the radius of the Sun and $M_{Sum}$ is the mass of the Sun, Scafetta (2012). The orbit of Mercury is elliptical and as the radius of Mercury, $R_M$, varies by 40% from 0.47 to 0.31 astronomical units (A.U.) during an orbit, the tidal effect varies by 120% during each orbit of 88 days. This is the largest percentage variation of tidal effect among the planets. In the following we specify the tidal effect of Mercury in terms of the quantity $(1/R_M)^3$ which varies from 9.8 A.U.$^{-3}$ to 34.4 A.U.$^{-3}$ during an orbit with average level 19.6 A.U.$^{-3}$.

The main reason for considering the effect of Mercury is illustrated in Figure 1 where the period of Mercury, 87.96926 days or ~ 88 days, the first sub harmonic, ~176 days, and the first harmonic, ~ 44 days, are seen to coincide almost exactly with the m = 7, m =14 and m = 3 Rossby mode periods respectively. The ratio between the period of the 1,3 mode, 41.917 days, and the first harmonic of the period of Mercury, 43.984 days, is 0.95; the ratio between the period of the 1,7 mode, 89.679 days, and the period of Mercury is 1.02; the ratio between the period of the 1,14 mode, 176.614 days, and the first sub harmonic period of Mercury, 175.938 days is 1.004; and the ratio between the period of the 1,28 mode, 351.85718 days, and the second sub harmonic period of Mercury, 351.87704 days, is 0.99994. Thus if the "additional mechanism promoting modes" is the tidal effect of Mercury one might expect to see evidence of Rossby wave modes on the Sun at periods of ~ 44 days, ~ 88 days, ~ 176 days and ~ 352 days. It is also useful to note that a Rossby wave moves around the surface of the Sun at an angular speed approximately proportional to $2/m^2$ times the angular speed of the Suns surface, Lou et al (2003). So for the higher m number modes the Rossby wave pattern is essentially stationary on the solar surface.

The concept that periodic activity on the Sun may be associated with the periodicity of the planets is controversial as the calculated tidal height variations on the Sun are minute. The tidal height difference due to Mercury varies from 0.60 mm to 0.17 mm during an orbit and is therefore insignificant compared with ambient fluctuations of the solar surface, De Jager and Versteegh (2005), Callebaut et al ( 2012), Scafetta (2012). The history of the planetary tidal effect concept has been reviewed by Charbonneau (2002). The predominant idea by proponents is that tides on the Sun due to the planets somehow stimulate or trigger solar activity and the period of this resultant solar activity is similar to the period of a single planet or to some more complex periodicity due to the conjunction or lining up of two or more planets, Hung (2007), Scafetta (2012), Tan and Cheng (2013). Scafetta (2012) has proposed that the tidal effect could be greatly amplified by a mechanism involving increased conversion of mass into energy according to Einstein's equation as the tidal effect causes minute density changes in the interior of the Sun. The variation of the tidal effect due to Jupiter is similar in amplitude to the variation due to Mercury. The period of Jupiter is 11.9 years and Scafetta (2012) has shown that, in combination with Earth and Venus, the combined tidal effect due to the three planets is phase coherent with the observed ~11 year solar cycle. The concept remains controversial due to the lack of a convincing mechanism that would allow minute tidal surface height variations to influence magnetic activity on the Sun with research continuing e.g. Wolff and Patrone (2010), Charbonneau (2013).

However, there is direct observational evidence of Rossby waves on the Sun with periods close to the periodicities associated with Mercury. Sturrock and Bertello (2010) provided a power spectrum analysis of 39,000 Mount Wilson solar diameter measurements between 1968 and 1997. They fitted Rossby wave mode frequencies to the eight strongest peaks in the spectrum. The two lowest frequency peaks in their tabulation were observed at 4.04 years$^{-1}$ and 2.01 years$^{-1}$. The corresponding periods, 90.4 days and 182 days, are each just 3% longer than, respectively, the period of Mercury and the period of the first sub harmonic of Mercury. Their observations therefore support the existence of Rossby wave modes on the Sun at periods very close to the periodicities associated with Mercury. It is also relevant that, in a comprehensive study of periodicity in sunspot area data covering solar cycles 12 to 21, Lean (1990) found the strongest peak in spectrograms of daily sunspot area occurred at 353 days, a period which corresponds almost exactly with the fourth sub harmonic of the period of Mercury, 351.88 days, and the n =1 m = 28 Rossby wave mode period, T(1,28) = 351.86 days, from equation 1.

Solar activity is associated with the emergence of magnetic flux on the surface of the Sun. This occurs, in an ~ 11 year cycle, when pockets of magnetic flux float up through the convection zone and emerge as groups of sunspots on the surface to the north and south of the Suns equator, Charbonneau (2010). In the absence of a periodic modulating mechanism, the emergence of sunspot groups should be completely random, Lou et al (2003). However, Rossby waves may trigger the onset of magnetic buoyancy instabilities, possibly of the Rayleigh-Taylor type, that lead to the periodic emergence of sunspots thereby linking the periodicity in sunspot number and sunspot area with the periodicity of equatorially trapped Rossby waves, Lou et al (2003).

To summarise: Mercury exerts a minute tidal force on the Sun with period ~88 days, Scafetta (2012). Theoretical estimates of Rossby wave periods predict allowed modes with periods very close to the periodicities associated with Mercury, Lou (2000). There is direct observational evidence of Rossby waves on the Sun with periods very close to the period and the first sub harmonic period of Mercury, Sturrock and Bertello (2010). Rossby waves may trigger the emergence of sunspots on the surface of the Sun, Lou et al (2003). A mechanism which practically promotes specific modes is required, Dimitropoulou et al (2008).

Bigg (1967), using techniques from radioastronomy for detecting periodic signals buried in noise, showed that daily sunspot numbers for the years 1850 to 1960 contained a small but consistent periodicity at the period of Mercury which is partially modulated by the positions of Venus, Earth and Jupiter. To our knowledge there are no other reported investigations of periodicity in sunspots due to Mercury. However, as indicated above, since 1967 evidence that Mercury may influence sunspots has increased with the development of the theory of solar magnetic Rossby waves, observation of Rossby wave modes at periods associated with Mercury and a plausible scenario linking Rossby waves and sunspot emergence. Therefore this paper re-examines the possibility that sunspots vary at periodicities associated with Mercury.

We use spectral analysis based on the Fast Fourier Transform (FFT) method to detect periodicities. This is the primary method of analysis used by other researchers to detect the Reiger type "quasi-periodicities" in the very large range of solar, space and terrestrial variables mentioned earlier. However, we will show that spectral analysis can lead to ambiguous results with, in many cases, the detected periodicities actually being related to the periods of sidebands rather than the period of the driving mechanism. For this reason we filter the sunspot data with band pass filters centred on 88 days and 176 days and compare the variation of the filtered data with the variation of the tidal effect due to Mercury. The purpose of the comparison is to demonstrate phase coherence between the variation of the filtered sunspot area data and the tidal effect of Mercury. The phase of the tidal effect of Mercury is known precisely at any time. Therefore, if sunspot area is related to the tidal effect of Mercury, it should be possible to demonstrate phase coherence of the band passed data and the tidal effect extending over intervals of tens or hundreds of years.

The arrangement of the paper is as follows: Section 2 outlines the method of spectral analysis and band pass filtering. Section 3 describes the result of band pass filtering the sunspot area data between 1876 and 2012. Section 4 examines phase coherence between the tidal effect of Mercury and the ~ 88 day and ~ 176 day variations in the sunspot area data during solar cycles 16, 18 and 23. Section 5 provides spectral analysis of the sunspot area data. Section 6 provides an estimate of the sensitivity of sunspot area to the tidal effect of Mercury during solar cycle 23 and Section 7 briefly examines the influence of Mercury on flares, solar wind speed and cosmic ray flux. Section 8 is the Conclusion.

**2. Method and data sources**
In the present study there are two main variables: the orbital radius of Mercury, $R_M$, and the sunspot area on the northern hemisphere of the Sun (SSAN). Daily sunspot area North (SSAN) is measured by the Greenwich Observatory in units of the area of one millionth of a solar hemisphere or microhems.

The data is available at http://solarscience.msfc.nasa.gov/greenwch/daily_area.txt

Daily values of the orbital radius of Mercury in A.U. are available at http://omniweb.gsfc.nasa.gov/coho/helios/planet.html for 1959 to 2019. Future or past values or the orbital radius of Mercury may be calculated using

$R_M = 0.38725 - 0.07975\cos[2\pi(t-24.5)/87.96926]$    A.U. (2)

where time in days, t, is measured from January 01, 1995.

The method of band pass filtering to obtain the ~ 88 day and ~176 day components of sunspot area is as follows. A Fast Fourier Transform of the entire sunspot area data series was made. The resulting n = 75 Fourier amplitude and phase pairs, $A_n(f_n)$, $\phi_n(f_n)$, in the frequency range 0.0102 days$^{-1}$ to 0.0125 days$^{-1}$ (period range 98 to 80 days) were then used to synthesize the ~ 88 day period band pass filtered version of sunspot area, denoted 88SSAN, by summing the 75 terms, $SSAN_n = A_n\cos(2\pi f_n t - \phi_n)$ for each day in the series. Similarly, Fourier pairs in the frequency range between 0.00511 days$^{-1}$ and 0.00625 days$^{-1}$ (period range 195 days to 160 days) were used to synthesize the ~ 176 day band pass filtered version of sunspot area denoted 176SSAN. Where data has been smoothed by, for example, a 365 day running average, the resulting smoothed data is denoted by the suffix Snnn e.g. a 365 day running average of sunspot area North data would be denoted SSAN S365.

**3. The ~ 88 day and ~ 176 day components of sunspot area North, 1876 to 2012.**

The amplitude of the ~88 day and ~ 176 day filtered components of sunspot area North are shown in Figure 2. The ~ 176 day component is displaced by 400 microhem for clarity. Also shown, the 365 day running average of the daily sunspot area North with numbers indicating the solar cycle. It is clear from the variation of amplitude of the filtered components in Figure 2 that the ~ 88 day and ~ 176 day components occur in "activations". The shortest activations last about 1.5 years e.g. the four activations of the ~88 day component during cycle 23. The longest activations last about 7 years e.g. the ~176 day component during cycles 23. Occasionally the activations are discrete as in cycle 23. However, more often activations overlap e.g. the three ~88 day activations in cycle 20. We will show in the next section that, when an ~88 day activation is discrete, the variation of the ~88 day component of sunspot area during the activation is phase coherent with the variation of the tidal effect of Mercury or, more simply, phase coherent with the orbital radius of Mercury. Similarly, when ~ 176 day activation is discrete the ~ 176 day variation, during the activation, is phase coherent with the first sub harmonic of the tidal effect of Mercury.

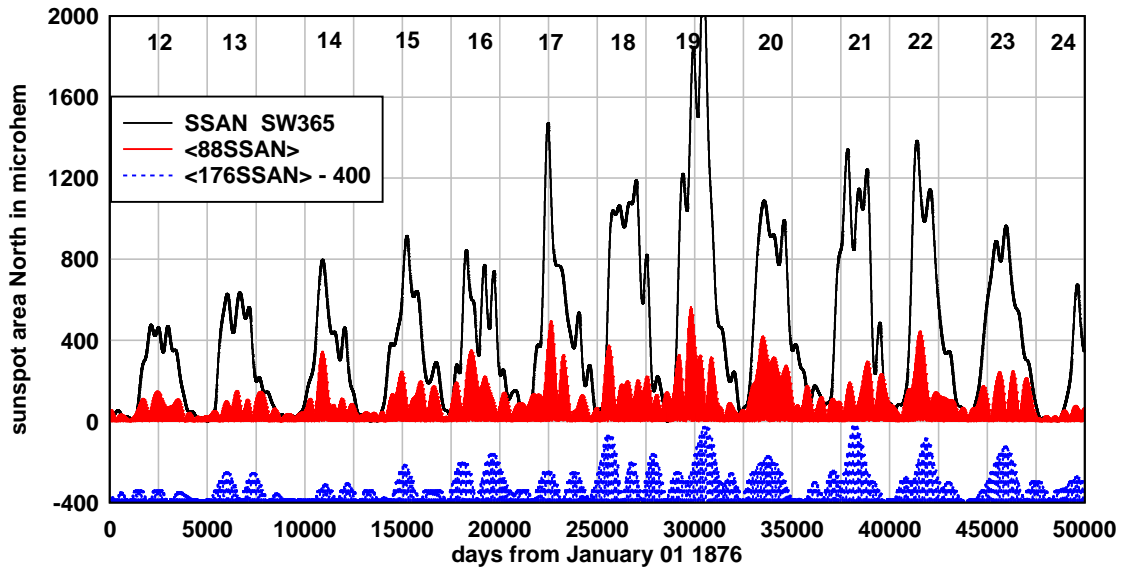

**Figure 2.** The amplitude of the ~ 88 day period (red) and ~ 176 day period (blue) components of sunspot area North from 1864 to 2012. Also shown the 365 day running average of sunspot area North, (black).

We infer that the "activations" shown in Figure 2 correspond to intervals of eruption of magnetic flux onto the surface of the Sun in the form of sunspots. Following Lou et al (2003), we suggest that Rossby waves at 88 days period and 176 days period trigger buoyant instability of sub surface magnetic flux resulting in pockets of buoyant magnetic flux periodically floating up through the convection zone and emerging as sunspots in the photosphere. This process continues in a dynamic feedback cycle similar to that described by You et al (2003) until buoyant magnetic flux in that region of the Sun is exhausted and the "activation" ceases. Sometimes only one region of the Sun is influenced by this process and the observed "activation" is discrete. At other times two Rossby wave modes are involved and two or more regions of the Sun are simultaneously "activated" and the activations overlap in time and in space.

It is clear from Figure 2 that the peaks in amplitude of the ~ 88 day and ~ 176 day components correlate with the pattern of peaks at the top of the sunspot area cycle. For example the twin peak pattern at the top of cycle 22 is clearly due to a strong ~88 day activation followed by a strong ~176 day activation. Similarly, the very square top form of solar cycle 18 is evidently due to five near equal amplitude and short ~88 day activations in combination with three near equal amplitude and discrete ~176 day activations. We infer from the results in Figure 2 that cycle to cycle changes in the shape of the sunspot cycle are simply the result of the combined effect of the separate ~88 day and ~176 day "activations" of sunspots during a cycle.

## 4. Phase coherence of sunspot area with Mercury during cycles 16, 18 and 23.

Figure 3 shows the ~ 88 day component of sunspot area North during 11 years of solar cycle 23. Also shown is the variation of $(1/R_M)^3$ from its average value, denoted $\Delta(1/R_M)^3$.

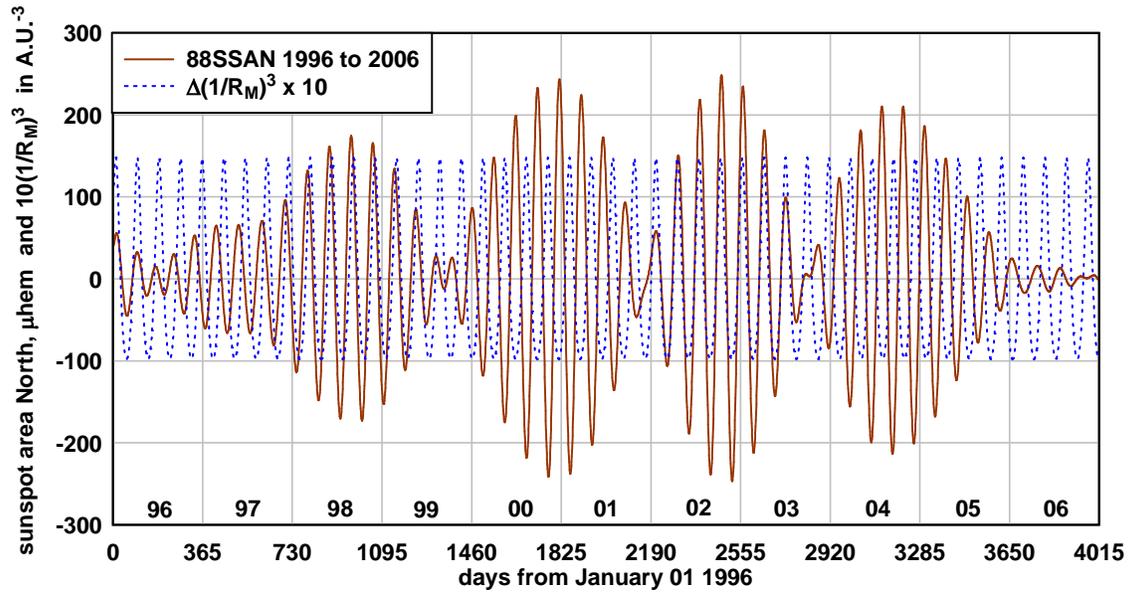

**Figure 3.** The ~ 88 day component of sunspot area North for 1996 to 2006 in solar cycle 23 (full line) compared with the variation of the tidal effect of Mercury. One weak and four strong activations are evident.

There are five activations of the ~ 88 day component of SSAN during cycle 23. A weak activation in 1997 overlaps the first strong activation in 1998. The strong activation in 1998 is followed by three strong and discrete activations; the first in 2000 2001, the second in 2002 2003 and the third in 2004 2005. During the four strong activations it is evident that the phase coherence between the ~ 88 day variation of sunspot area North and the tidal effect of Mercury alternates between exact in-phase coherence to exact anti-phase coherence. We infer from this result that when activations are discrete succeeding activations usually occur with a π phase shift relative to the preceding activation. Why this π phase shift occurs is not clear. However, as the allowed periods of 1,7 and 2,7 Rossby wave modes are nearly the same, see Figure 1, it may be that successive activations respond alternately to n = 1 or n = 2 Rossby wave modes and this may account for the observed π phase shift. However, we note that, taking account of the π phase shifts between the alternating discrete activations, the ~ 88 day component of sunspot area North is exactly phase coherent with the tidal effect of Mercury from 1998 to 2005 or over 33 orbits of Mercury.

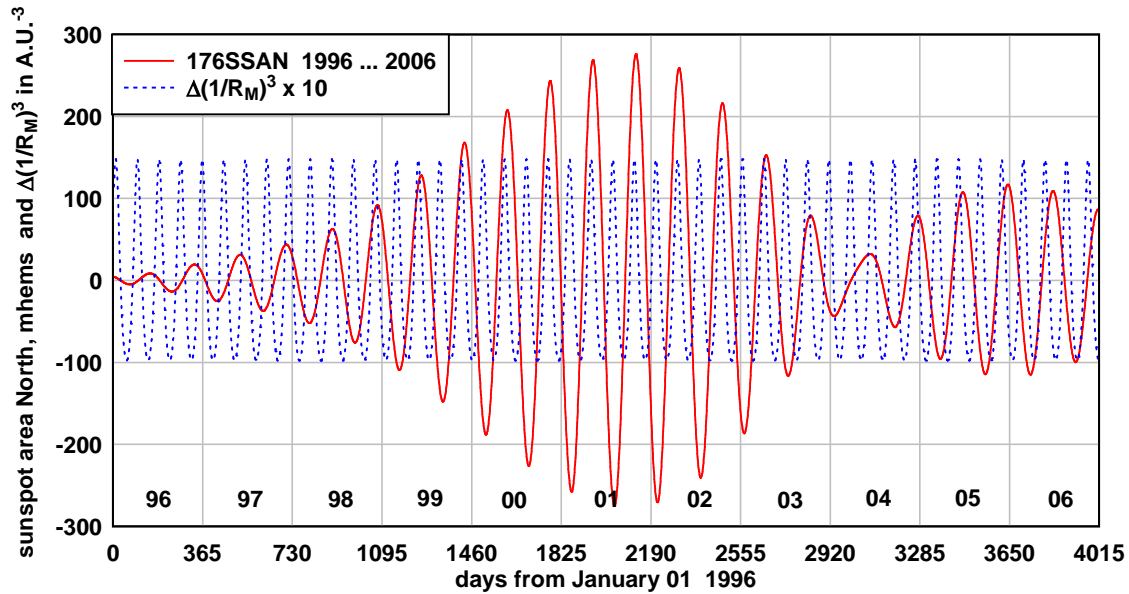

**Figure 4.** The ~ 176 day component of sunspot area North for 1996 to 2006 in solar cycle 23 (full line) compared with the variation of the tidal effect of Mercury. There is one dominant activation during this interval.

Figure 4 shows the ~ 176 day component of sunspot area North and $\Delta(1/R_M)^3$ during 11 years of solar cycle 23. There is a strong, discrete activation extending from about 1997 into 2003. This is followed by a weaker activation beginning in 2004. Thus the ~176 day component in cycle 23 is dominated by the first activation. The first activation is clearly in phase with every second peak or the first sub-harmonic of the tidal effect of Mercury for the seven years from 1997 to 2003 or over 27 orbits of Mercury. Careful examination of Figure 4 shows that in the second activation, beginning in 2004, the phase of the variation is, initially, shifted by one orbital period or 88 days from the variation in the first activation, i.e. the variation in the second activation begins with a π phase shift relative to the variation in the first activation. However, we note that, by 2006, the phase of the variation in the second activation appears to be moving by π/2 to a position where the peak of the ~176 day sunspot area response lies between every second pair of peaks of the Mercury tidal effect. To investigate the latter effect more closely we examine the ~176 day variation during solar cycle 16.

From Figure 2 we know that there are two strong, discrete, ~176 day activations during cycle 16. Figure 5A compares the ~176 day component of sunspot area North with the tidal effect variation of Mercury during solar cycle 16. During the first activation from 1923 to 1927 the ~176 day component is exactly phase coherent with every second peak of the tidal effect of Mercury. During the second activation the ~176 day variation settles, during 1929 and 1930 to be phase coherent with each successive pair of peaks of the tidal effect or in anti-phase with every second minimum of the tidal effect.. This implies that the phase shift between the first and second activations, while initially close to π, settles, after one or two 176 day cycles, to a phase shift of 3π/4. This is more clearly seen in Figure 5B where the time axis is divided into 176 day intervals. Again it is not clear why this phase shift between sequential discrete activations occurs. As discussed above, it may be due to a change from response to the 1,14 wave mode to response to the 2,14 wave mode. In any case, taking into account the systematic phase change of $3\pi/4$, during the eight years between 1923 and 1930 the ~176 day component of sunspot area North is phase coherent with the tidal effect of Mercury.

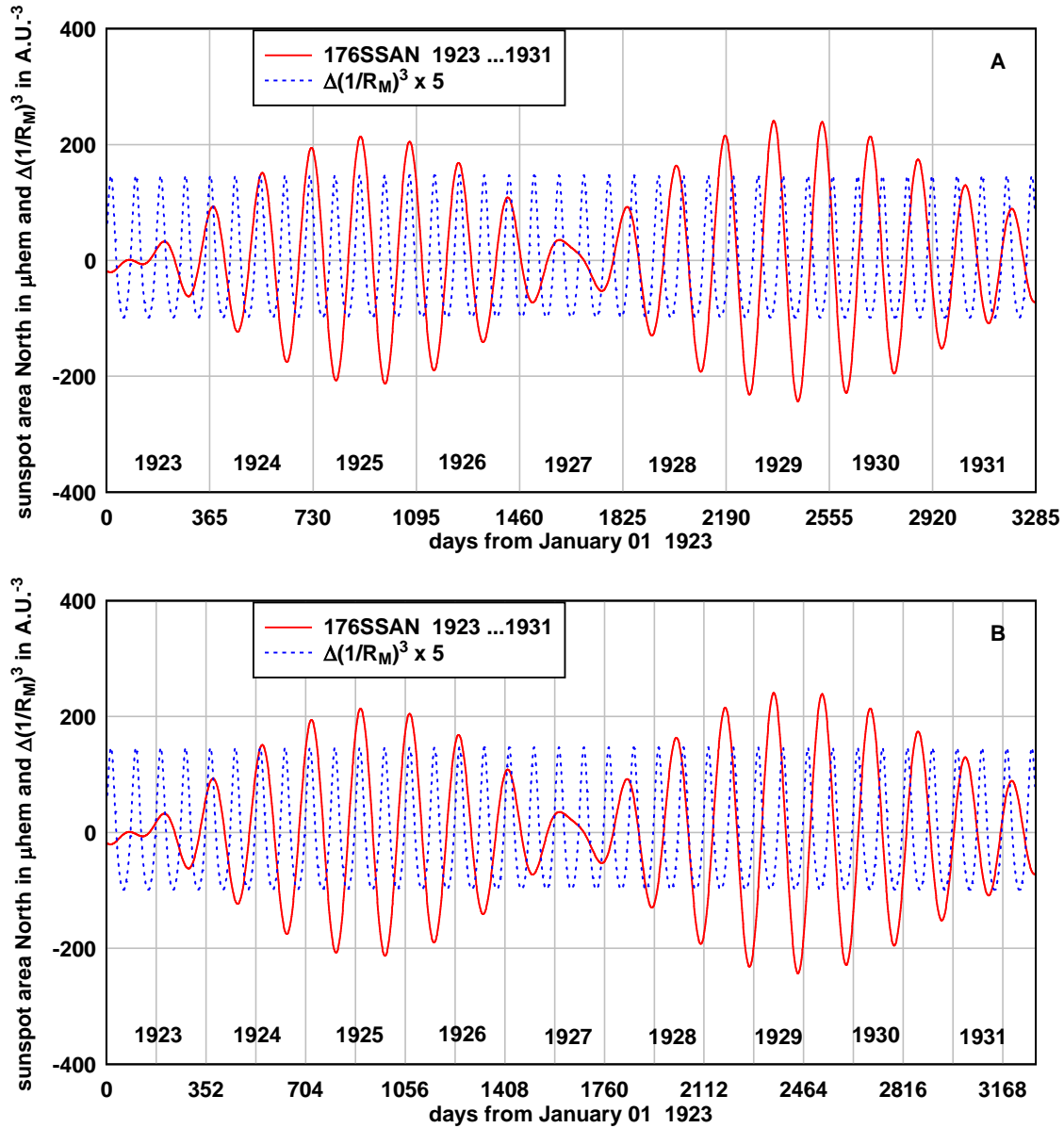

**Figure 5.** The ~ 176 day period component of sunspot area North from 1923 to 1931 during solar cycle 16 showing two discrete activations. In 5A the time axis is in years, in 5B the time axis is in intervals of 176 days to indicate more clearly the phase change in the ~ 176 day period variation.

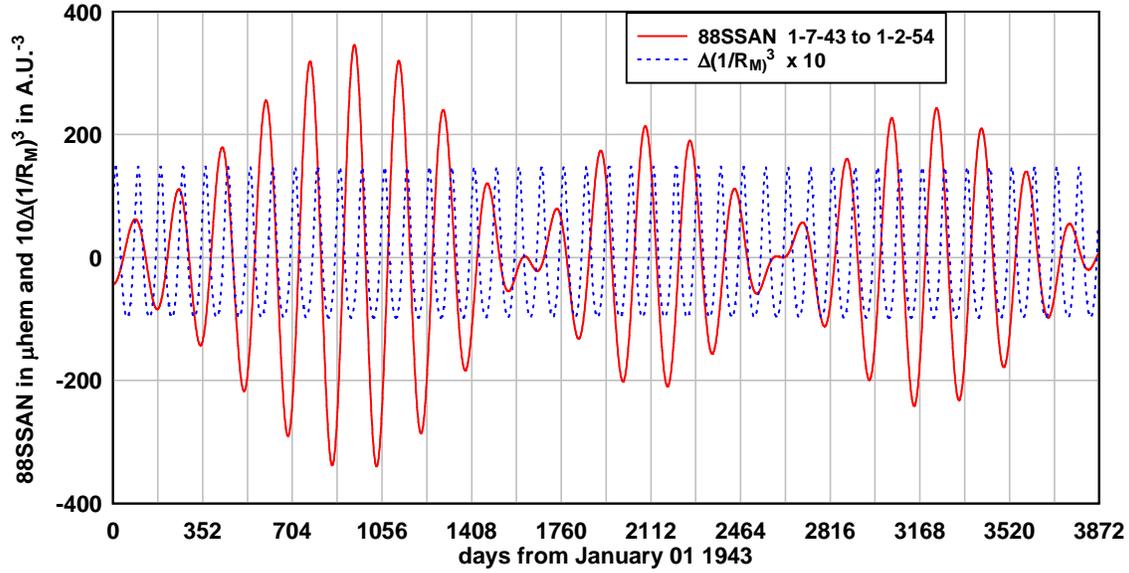

**Figure 6.** The ~ 176 day period component of sunspot area North from 1943 to 1954 during solar cycle 18 showing three discrete activations. The time axis is in intervals of 176 days to indicate more clearly the phase change in the ~ 176 day period variation.

Figure 6 shows the ~176 day variation during cycle 18 as a third example. Here the peaks of the ~176 day first sub harmonic response in all three activations occur between sequential pairs of peaks of the Mercury tidal effect. However, with the time axis divided into intervals of 176 days, it is evident that there is a π phase change of the ~ 176 day response between the first and second activation and between the second and third activation. In a nonlinear system there will always be a stable harmonic response at frequencies that are multiples of the driving frequency. However, sub harmonic response is usually associated with some instability in the system and usually occurs only when some threshold level of the driving source is exceeded, Yen (1971). We assume that, in the present case, the instability is associated with buoyant instability of magnetic flux.

In most solar cycles there is considerable overlap between activations. For example Figure 2 shows that in cycle 20 there is strong overlap of three ~88 day activations. When this is the case it is not possible to discern the clear phase coherence that is obvious in the case of discrete activations as illustrated above. Further it is difficult to ascertain whether sequential, overlapping, activations suffer π phase change. Reference to Figure 2 shows that solar cycle 14 illustrates a case where there are both discrete and overlapping activations in the ~ 88 day component of sunspot area North, reproduced in Figure 7. The weak activation in 1903 1904 is discrete from the strong activation in 1905 1906 which itself overlaps with a weak activation in 1907. The first activation is in phase with the tidal effect and the second, strong activation begins, with a π phase change to be in exact anti-phase with the tidal effect. However, towards the end of 1906 this activation overlaps with the third, weak, activation and the anti-phase variation with the tidal effect becomes less exact. It is interesting to note that because the phase of the tidal effect of Mercury is known precisely we can infer that, when activations are discrete, the ~88 day and ~176 day components of sunspot area are phase coherent with the variations in other discrete activations occurring in the past or future. For example, we know the ~88 day period variations in sunspot area between 1903 and 1904 in Figure 7 must be exactly phase coherent with the ~88 day period variations in sunspot area between 2002 and 2003, Figure 3, simply because the Mercury tidal effect is a precisely constant period time reference.

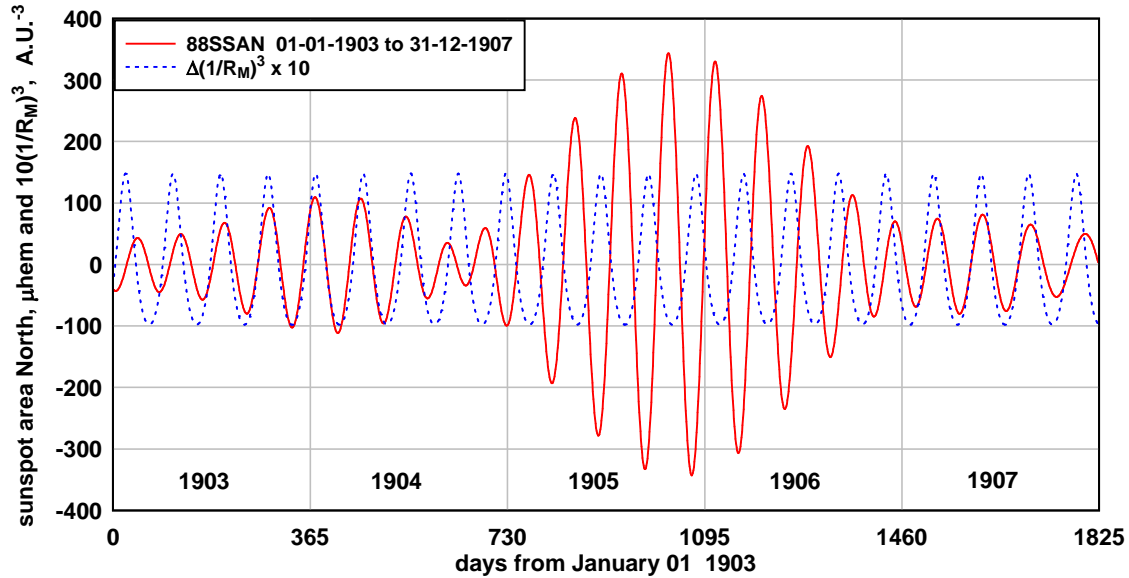

**Figure 7.** The ~ 88 day component of sunspot area North for 1903 to 1907 in solar cycle 14 (full line) compared with the variation of the tidal effect of Mercury. One weak and four strong activations are evident.

It is interesting to note that Lean (1990) investigated the phase coherence of the 155 day "quasi-periodicity" in successive solar cycles using a method of fitting sinusoids to sunspot area data band pass filtered with the band pass centred on 155 days. Lean (1990) found that "While these investigations of the 155 day periodicity do indicate approximate phase coherence between adjacent solar cycles, they also illustrate that the 155 day periodicity is not equivalent to a deterministic, sinusoidal oscillation at one fixed period. Rather, the actual period drifts with time, and uncertainties in knowing this period propagate as errors when the cycle is extrapolated too far beyond the time interval in which the actual period has been determined." We will show, in the next section, that the ~155 day quasi-periodicities are sidebands of the 176 day first sub harmonic of the Mercury tidal effect and that "quasi-periodicities" obtained from a spectrogram depend on three parameters; the driving periodicity or its sub harmonics, the number of activations encompassed by the spectrogram and the average time interval between activations. Lean (1990) also determined, by band pass filtering centred on 155 days, the amplitude of the ~155 day component in sunspot area North for solar cycles 12 to 21 finding a similar time and amplitude variation of activations or "episodes" as we report in Figure 2. This is expected as the amplitude of any sideband component is expected to vary proportionately to the amplitude of the driving component or the amplitude of its harmonics or sub harmonics. It is interesting to ponder the outcome had Lean (1990) used a filter centred on 176 days or 352 days rather than one centred on 155 days.

## 5. Spectral analysis of the sunspot area North data.

Since the discovery of "quasi-periodicities " in solar flare data by Reiger (1984) the discovery and study of Reiger type "quasi-periodicities" in a wide range of space variables has been conducted, principally, by means of spectral analysis. Nearly all of the reports of "quasi-periodicities" in solar related variables referred to in the introduction are based on the discovery of peaks in the 50 day to 200 day period range in spectrograms or time-frequency diagrams of variables ranging from sunspot number to cosmic ray flux. Figure 2 showed that the solar activity in this period range occurs in activations lasting from about 1 to 7 years. It follows that any variation in solar activity or solar related variable will be amplitude modulated by this type of activation. The spectrum of an amplitude modulated signal contains sidebands and this introduces an ambiguity in assessing the origin of a particular periodicity. It is clear that before assessing a spectrogram of a variable like sunspot area in, for example, cycle 23 where sequential discrete activation occurs it is necessary to understand the amplitude modulation effect of the activations on a spectrogram. So, as a preliminary to this section, we study the simple amplitude modulation,

$$y = [A+\sin(2\pi t/730)]\sin(2\pi t/88) = [A+\sin(2\pi f_m t)]\sin(2\pi f_1 t). \qquad (3)$$

The term in square brackets is a two year period modulation and this term modulates the 88 day period sinusoid. When $A \gg 1$ there is no modulation and a spectrogram of y contains a single peak at 88 days. When $A \sim 1$ there is strong amplitude modulation and side bands appear at $f_S = f_1 +/- f_m$ in the spectrogram, Figure 8A and 8B.

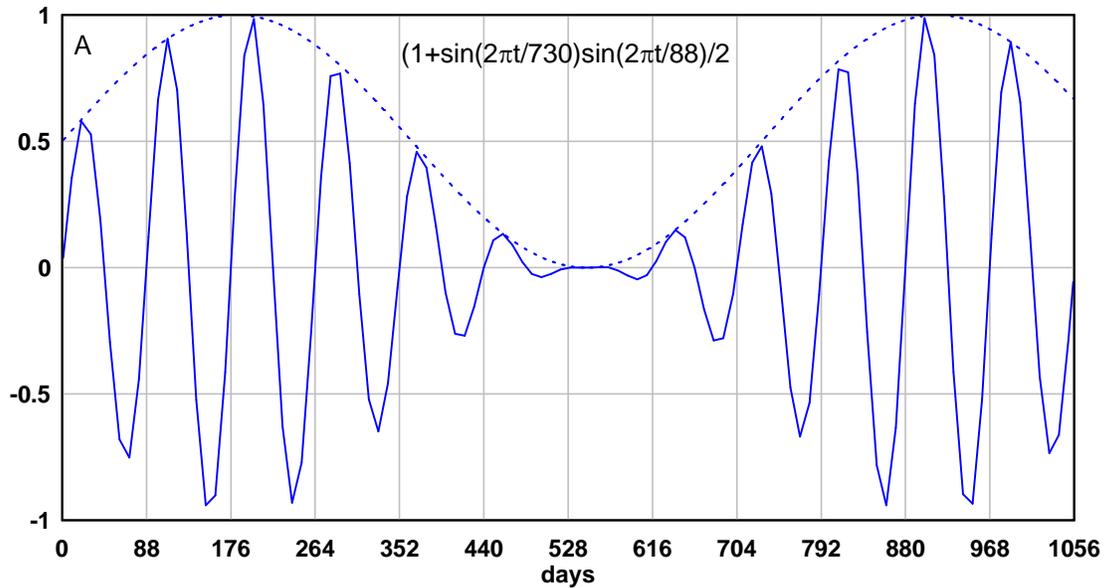

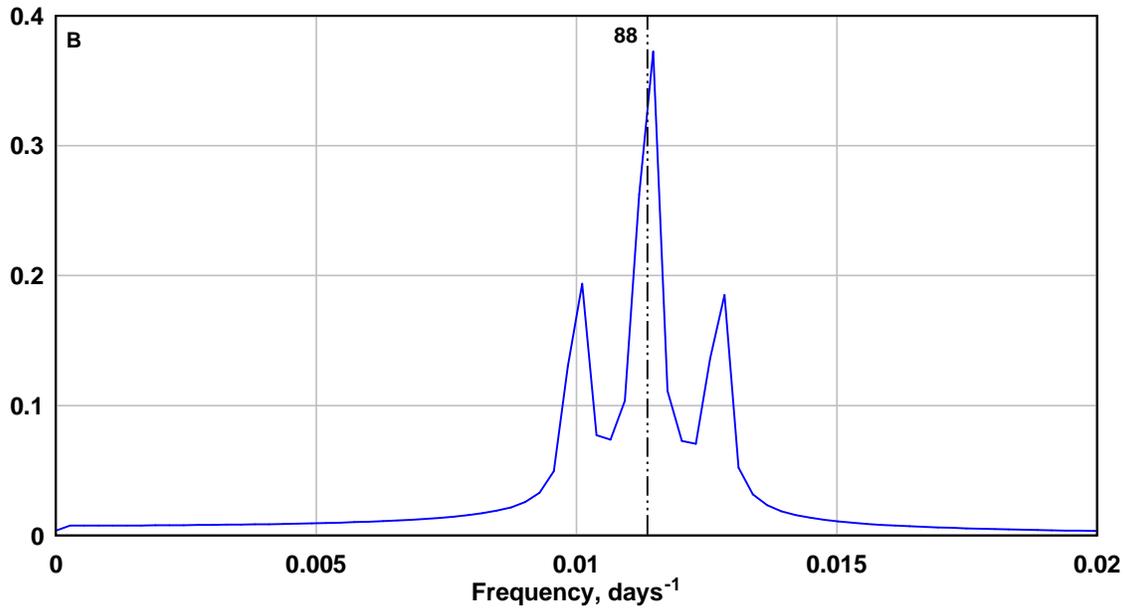

**Figure 8.** The variation (A) and the spectrogram of the variation (B) of an amplitude modulated sinusoid when A = 1 in equation 3. The time axis of the variation is in intervals of 88 days so that any phase change can be followed. In this case there is no phase change from one modulation maximum to the next.

When A << 1 the modulation is said to "cross zero", the sign of the 88 day sinusoid is reversed when the modulating term becomes negative, as a result there is a π phase shift in the 88 day signal between one modulation peak and the next, and nearly all power is shifted from the central peak at $f_1$ to the two sideband peaks at $f_1$ +/- $f_m$, Figure 9A and 9B.

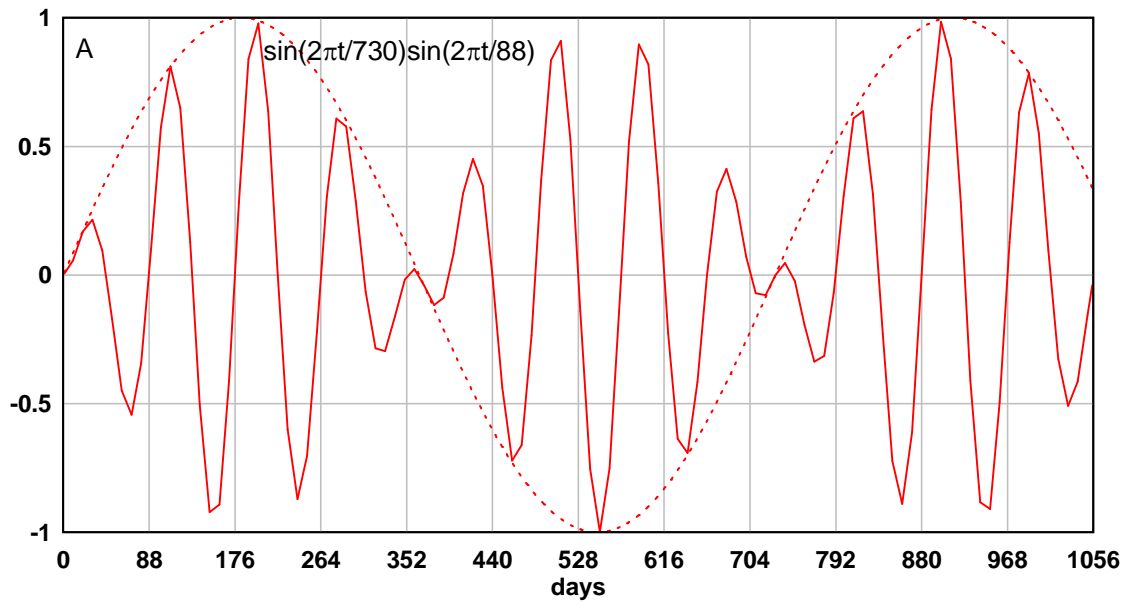

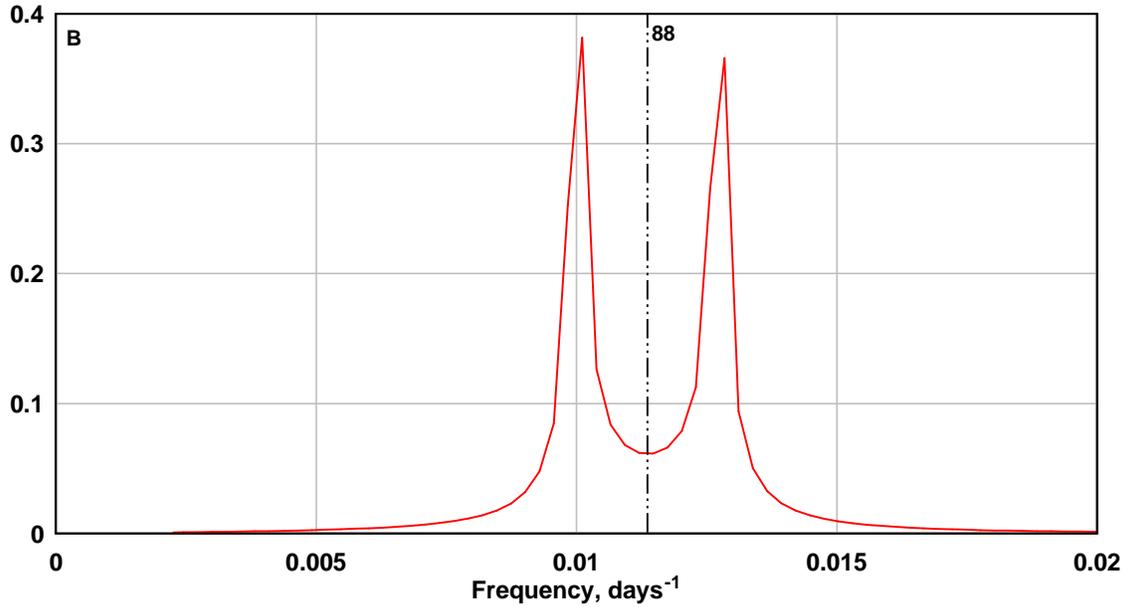

**Figure 9.** The variation (A) and the spectrogram of the variation (B) of an amplitude modulated sinusoid when A = 0 in equation 3. The time axis of the variation is in intervals of 88 days so that any phase change can be followed. In this case there is a π phase change from one modulation maximum to the next.

The time variations in Figure 8A and 9A are shown over 1056 days for clarity. However the spectrograms or Fast Fourier Transforms, Figures 8B and 9B, are calculated over 3650 days to provide better resolution. It is evident from the simulations above that the variation during any one modulation maximum or "activation" is at 88 day period and a spectrogram made over a single modulation maximum or activation will have the major peak at 88 day period. However, if a spectrogram is made over several activations when A ~ 0 the spectrogram may show no peak at 88 days and only sidebands on either side of a central minimum as in Figure 9B.

The simple modulation relation of equation 2 is a useful means of interpreting the effect of the activations of sunspot area observed in Figures 2, 3, 4, 5, 6 and 7. For example, in the sunspot area North data for solar cycle 23, Figure 3, four strong activations occur, separated on average by ~1.5 years or ~ 550 days with a π phase shift between the variations in succeeding activations. Thus, A = 0, $T_1$ = 88 days, $T_m$ = 1100 days in equation 2 would simulate this case. The expected spectrogram should evidence a minimum at $T_1$ = 88 days or $f_1$ = 0.0114 days$^{-1}$ and strong sidebands at $f_1$ -/+ $f_m$ i.e. at frequencies 0.0104 days$^{-1}$ and 0.0123 days$^{-1}$. The corresponding sideband periods are 81 days and 96 days.

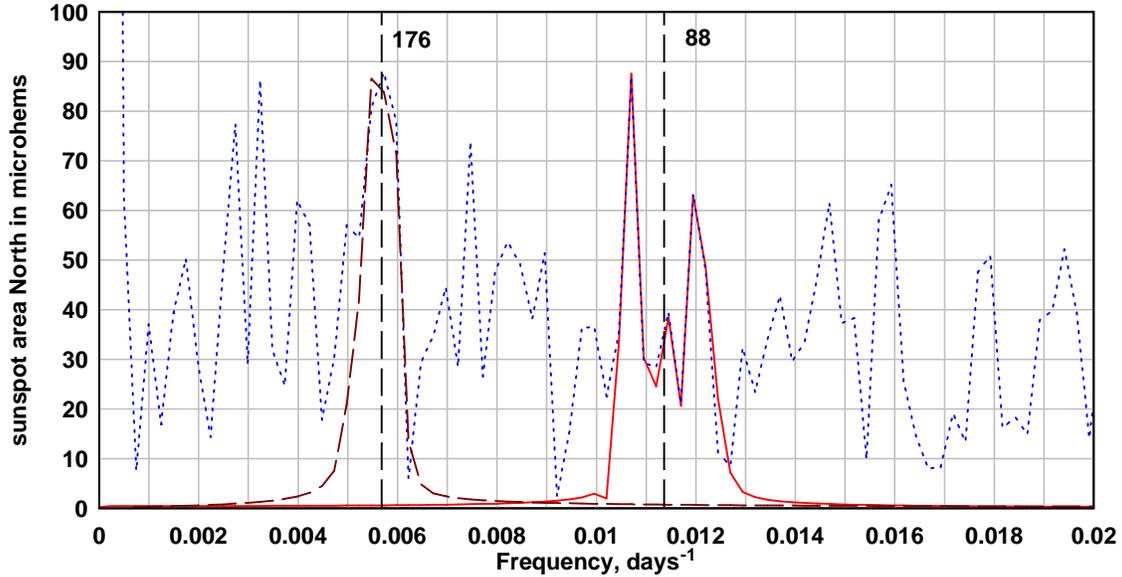

**Figure 10.** The spectrogram of daily sunspot area North during the interval 1996 to 2006 in solar cycle 23, (blue dots). The spectrograms of the ~ 88 day and ~ 176 day components are superimposed.

The observed spectrogram of the sunspot area North data for the eleven years between 1996 and 2006 is shown as the blue dotted line in Figure 10. The spectrograms of the band pass filtered data, the ~ 88 day period variation and the ~ 176 day period variation, are also shown in Figure 10. The spectrum around the 88 day period shows the two sidebands on either side of a minimum at 88 day period expected when the modulation crosses zero and the 88 day period variations in alternating modulation peaks differ by $\pi$ in phase. As indicated above this situation is expected for $A \sim 0$ in equation 3. The spectrum also shows a strong peak at 176 days. From the simulations outlined above this is characteristic of relatively weak modulation i.e. $A > 1$ in equation 3. We know from Figure 2 and Figure 4 that there is a long, strong, and discrete activation of the ~176 day component of sunspot area North in solar cycle 23. This is followed by a short, weak activation so the spectrogram is dominated by the effect of the first activation. Thus the spectrograms of Figure 10 are consistent with the type of activation observed in cycle 23 for the ~176 day component as well as for the ~88 day component of the sunspot area data.

An example of a solar cycle that is dominated by a single activation of the ~176 day component and a single activation of the ~88 day component is, from Figure 2, solar cycle 22. In this cycle there is a dominant ~ 88 day activation and a dominant ~176 day activation and we would, therefore, expect a spectrogram of sunspot area to show a strong peak at ~88 day period and a strong peak at ~176 day period and relatively weak sidebands. Using the standard FFT method we find that this is the case. However, we note that Oliver and Ballester (1995) using the more powerful methods of Lomb-Scargle periodogram and the Maximum Entropy Method found the most significant peak in sunspot area spectrum during cycle 22 was at 133 nHz (87 days). A peak at 66 nHz (175 days) was also found in most of the intervals in cycle 22 investigated by Oliver and Ballester (1995). However, no peak near 74 nHz (155 days) was found. We see the results of Oliver and Ballester (1995) as confirmation of the inference above that "quasi-periodicities" other than 88 day and 176 day period should be interpreted as sidebands associated with the 88 day period or 176 day sub harmonic period of the tidal effect.

Solar cycle 23 was chosen to illustrate the effect activations have on the spectra of sunspot activity because the activations of the ~88 day component and of the ~ 176 day component during solar cycle 23 were discrete and relatively simple. When the activations overlap and/or are variable in strength e.g. cycle 19 and 20 in Figure 2, simulations similar to, but more detailed than, the simulation of equation 3 can still be useful but the situation is less simple to interpret.

The main objective of this section was to demonstrate the potential for drawing ambiguous conclusions from spectrograms when attempting to discover or study Rieger type quasi-periodicities in solar activity related data. It is clear from the above simulations and observations that a sideband of a modulated signal may appear as a peak in a spectrogram when a peak at the period of the driving mechanism is absent. For example if one wishes to discover ~160 day "quasi-periodicities" in solar activity such as sunspot area one can note that the required sideband frequency, $f_S$, is 0.00625 days$^{-1}$. Use of the sideband relation $f_S = f_1 +/- f_m$ with $f_1 = 1/176$ days$^{-1}$ indicates that the required modulation frequency is $f_m = 0.00057$ days$^{-1}$ or modulation at a period of 4.8 years. Thus intervals of discrete activations of the ~176 day component that are separated by ~ 2.4 years in Figure 2 will coincide with intervals when ~160 day "quasi-periodicities" are likely to be discovered. Solar cycle 18 is a good example to illustrate this effect as it has three discrete activations of ~176 day periodicity each separated by about 2.4 years. In Figure 11 the broken red line shows a simulation of the spectrogram expected when applying the "discovery" procedure outlined above and the full blue line is the actual spectrogram of the sunspot area North data between July 1 1944 and July 1 1953 in solar cycle 18. Evidently a ~160 day "quasi-periodicity" has been "discovered" in sunspot area North during this interval whereas the true driving periodicity is the 176 day first sub harmonic of the period of Mercury. We note again that the period of a sideband peak depends on three parameters; the driving periodicity or periodicity of its sub harmonics, the number of activations encompassed by the spectrogram and the average time interval between activations.

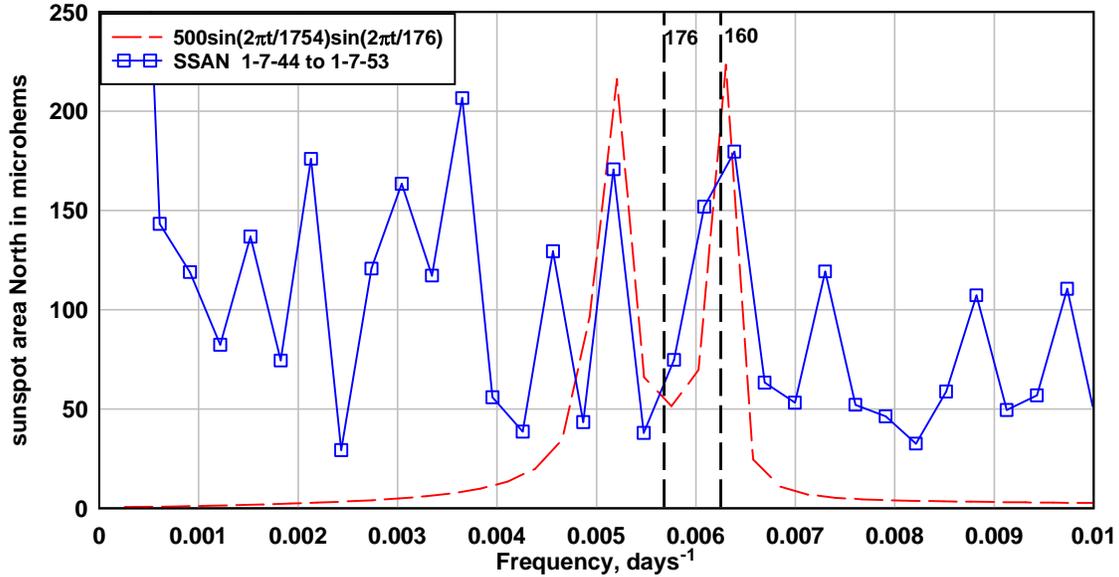

**Figure 11.** Illustrates the method for "discovering" ~ 160 day periodicities in sunspot area data. The broken red line shows the spectrogram due to the variation of equation 3 with $A = 0$, $f_m = 0.00057$ days$^{-1}$ and $f_1 = 1/176$ days$^{-1}$ and the blue squares show the actual spectrogram between 1944 and 1953 in solar cycle 18.

The results of Section 4 indicated that a $\pi$ phase shift in the variation between sequential activations is the usual case. This Section shows that spectrograms taken over several discrete activations have a minimum at the driving period with sidebands on either side, the periods of which depend on the duration of activations. We infer from this that it may be more useful to search for deep minima rather than high peaks in the spectrograms of variables connected with solar activity. We also infer that spectrograms obtained in previous research could be interpreted from this point of view. A good example is the spectrum of daily counts for flares in cycles 20 and 21, Bogart and Bai (1985), Figure 2C, where a deep minimum at 176 days (0.065 μHz) has several closely spaced sidebands on either side of the minimum suggesting several activations of different duration. contributed during cycles 20 and 21 as well as modulation by the solar cycles.

**6. Sensitivity of sunspot area to the tidal effect of Mercury during solar cycle 23.**

In Figure 12A we show the variation of daily sunspot area North for the 11 year interval from 1996 to 2006 during cycle 23. Also shown is the 365 day running average of this data which gives an indication of the variation due to the ~11 year cycle. The lower two curves are the ~ 88 day period and the ~ 176 day period components of sunspot area North. When the two components are added to the 365 day running average the full blue curve shown in Figure 12A is obtained. The variation of the blue curve provides an indication of the contribution of the ~88 day and ~176 day components to the overall variation. Figure 12B shows the sunspot area North variation after the 365 day moving average is subtracted, (grey dots). The standard deviation of this variation is 453 μhem. Also shown is the sum of the ~88 day period and ~176 day period components (blue line). The standard deviation of this combined variation is 144 μhem. Thus the components associated with the period and first sub harmonic period of Mercury contribute about 1/3 of the short term variation in sunspot area North i.e. about 1/3 of the variation at periods < 365 day period.

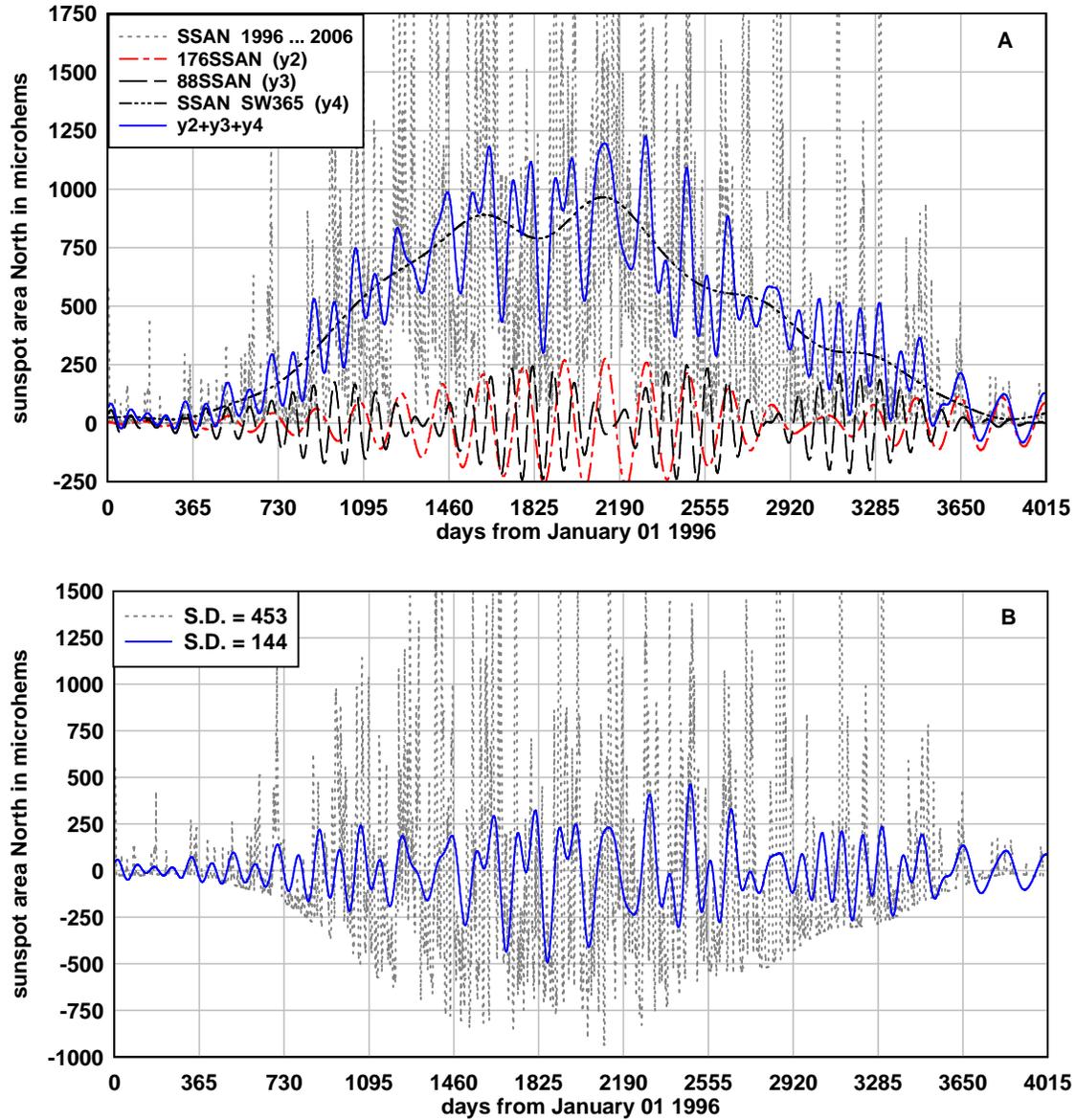

**Figure 12.** (A) The daily sunspot area for the 11 years from 1996 to 2006 in solar cycle 23 (grey dots) with the 365 day moving average superimposed (broken black). The full blue curve is the sum of the ~88 day component (black dash) and ~176 day component (red dash dot) with the 365 day moving average. (B) The daily sunspot area North in solar 23 with the 365 day moving average removed (grey dots) and the sum of the ~88 and ~176 day components (full blue curve).

The standard deviation of the variation in the tidal effect of Mercury, $\Delta(1/R_M)^3$, is 8.6 A.U.$^{-3}$. Thus, on average over cycle 23, a variation of 144 μhem in sunspot area North results from a variation of 8.6 units in the tidal effect. Normalising, each unit variation in $\Delta(1/R_M)^3$ results in 16.7 μhem variation in the combined effect of the ~88 day and ~176 day components. The mean level of the tidal effect, $(1/R_M)^3$, is 19.6 A.U.$^{-3}$ and the mean level of the sunspot area North variation during cycle 23 is 426 μhem, including the ~11 year variation. So a fraction 1/19.6 or 5% variation in tidal effect results in a 16.7/426 or 3.9% change in the average level of sunspot area North. As the tidal effect varies by 120% during each orbit of Mercury we find that each orbit contributes, on average during solar cycle 23, to a 94% change in the sunspot area.

To check how such a large sensitivity is obtained consider the variations near day 1825 in Figure 12A. At this time the average variation due to the ~88 day and ~176 day components associated with Mercury is about 500 μhem peak to peak. The average level of the sunspot area at day 1825 is about 750 μhem. Thus each orbital variation contributes a fractional change of about 500/750 or a 67% change in the average level of sunspot area. However, the percentage variation during the ascending and descending part of the cycle is much higher. For example at day 3285 during the descending phase the peak to peak variation due to the combined ~88 day and ~176 day components is about 400 μhem while the average sunspot area level is only about 250 μhem. Thus, at this time, the fraction contributed by the Mercury period components to the variation during one orbit is 400/250 or 160%.

Clearly a large fraction of the variation in daily sunspot area North during cycle 23 is due to components coherent with the variation of the tidal effect of Mercury and the first sub harmonic of the tidal effect. However, examination of Figure 12B shows that each maximum in the combined ~88 day and ~176 day variation (the full blue line) is accompanied by large fluctuations in sunspot area at shorter periods. Although this paper has not examined the shorter period variation it is possible that the shorter term variation evident in Figure 12 may be associated with harmonics, at 44, 22 and 11 days of the 88 day tidal effect and harmonics, at 58, 35 and 25 days of the 176 day first sub harmonic of the tidal effect of Mercury. Note, however, that some of the shorter period harmonics overlap the periodicity range due to solar rotation and would be difficult to separate. Also this paper does not examine the contribution from higher sub harmonics of the Mercury tidal effect. In particular periodicities at the sub harmonic periods of 352, 440 and 528 days are not assessed. It is worth recalling that Lean (1990) observed that the most significant peak in 100 years of sunspot area North data was at 353 days. The overall contribution of the tidal effect of Mercury to the variation of sunspot area is likely to be higher when the harmonics and higher sub harmonic components are considered.

**7. The connection to other space variables.**

**7.1 Sunspot number**
During solar maximum numerous sunspots and coronal holes emerge in the equatorial region and, in many cases, the sunspots and coronal holes appear to be spatially connected, e.g. Bravo et al (1998). The variation of space variables such as solar wind velocity, interplanetary magnetic field and cosmic ray flux depend, usually in a complex way, on the combined effect of magnetic flux associated with the polar and toroidal coronal magnetic fields as well as magnetic emanations from sunspots on both hemispheres of the Sun. If, as expected, the influence on space variables is due to the combined effect of activity in both hemispheres it is useful to examine how sunspot emergence differs in the two hemispheres. Figure 13 compares the ~88 day period component of sunspot area South (88SSAS) with the variation of the tidal effect of Mercury during cycle 23.

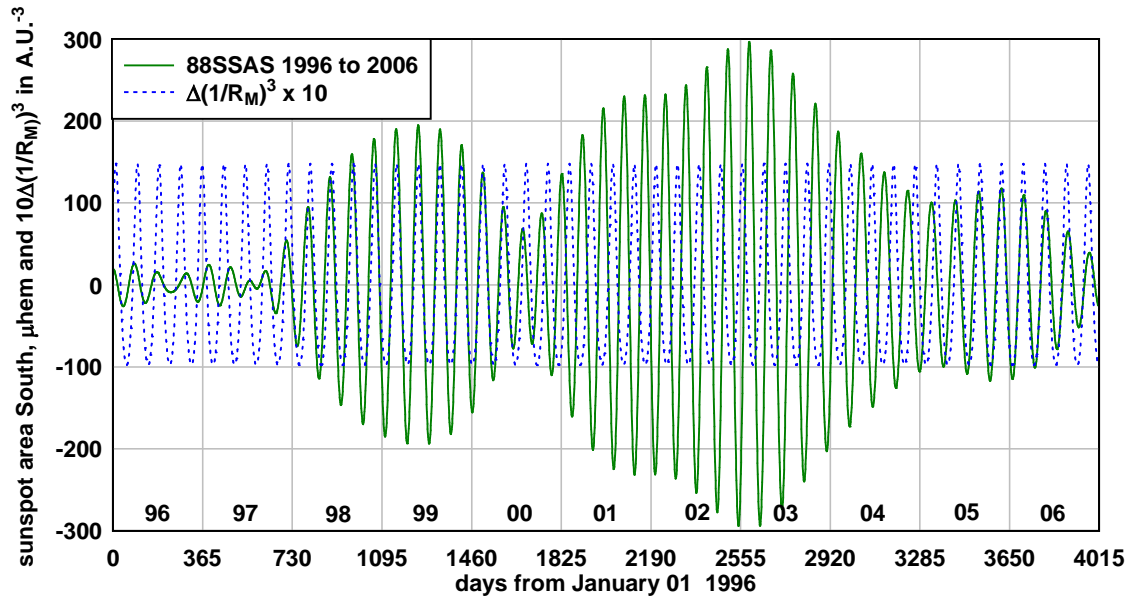

**Figure 13.** The ~ 88 day component of sunspot area South for 1996 to 2006 in solar cycle 23 (full line) compared with the variation of the tidal effect of Mercury. One discrete activation (1998 1999) is followed by three overlapping activations.

Figure 13 can be compared with Figure 3 which shows the variation of the ~ 88 day period component of sunspot area North during cycle 23. The two variations are quite different. In the Southern hemisphere there is a discrete activation of the ~88 day component during 1998 1999 which is, as expected, phase coherent with the Mercury tidal effect. This first activation is followed by three strongly overlapping activations where phase coherence with the Mercury tidal effect is less evident.

In Figure 14 we superimpose the variations of the Southern and the Northern hemisphere ~ 88 day components of sunspot area. During 1998 and extending into 1999 the two variations are in phase. We infer from Figure 14 that in phase Rossby wave modes exist in both hemispheres during 1998 1999 and that periodic and in phase sunspot emergence is triggered by the in-phase Rossby waves. Around mid 1999 the activation in the Northern hemisphere diminishes to near zero and, for the latter half of 1999, sunspot emergence is dominated by the Rossby wave mode in the Southern hemisphere and sunspots occur predominantly in the Southern hemisphere. For the remainder of solar cycle 23 there are intervals when both of the ~88 day variations are in phase or the variation in one hemisphere is dominant. However, it is clear that there is no long interval of phase coherence between the Mercury tidal effect and both sunspot area variations other than during the 1998 1999 interval.

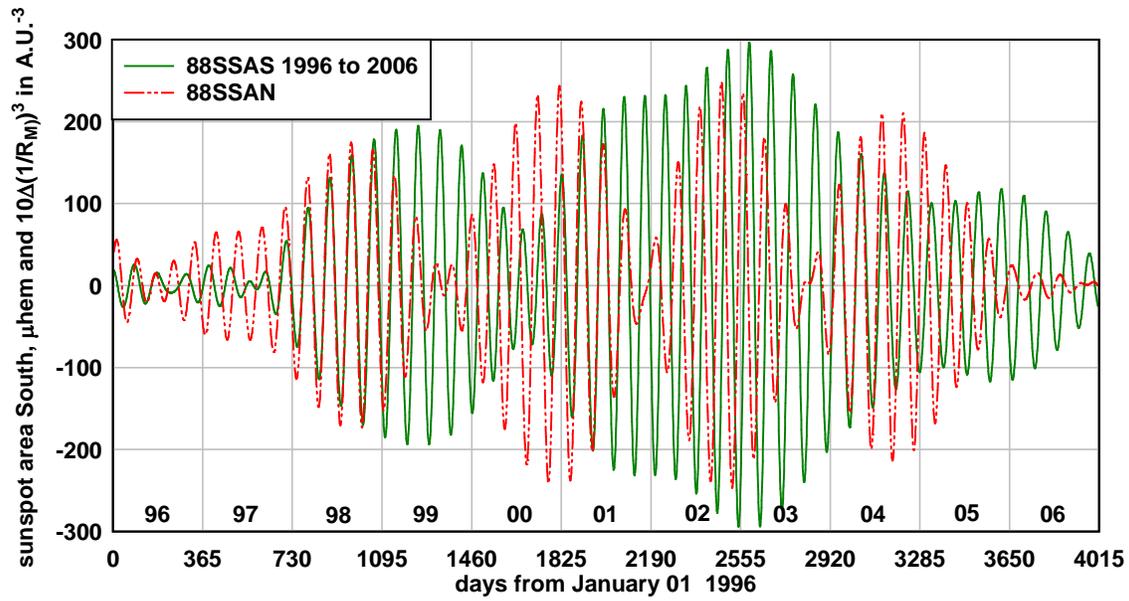

**Figure 14.** The ~ 88 day component of sunspot area South for 1996 to 2006 in solar cycle 23 (full green line) superimposed on the ~88 day component of sunspot area North for the same interval (dash dot dot red line).

Any variable that depends on activity in both solar hemispheres is expected to show exact phase coherence with the Mercury tidal effect only during the 1998 1999 interval. Daily sunspot number (SSN) is one of the numerous variables that depend on the combined effect of magnetic activity in both hemispheres. SSN has been measured daily from 1818 and 2013 and is available at:
http://www.ngdc.noaa.gov/stp/space-weather/solar-data/solar-indices/sunspot-numbers/international/listings/listing_international-sunspot-numbers_daily.txt.

Figure 15 compares the ~ 88 day period component of sunspot number (88SSN) with the variation of the tidal effect, $\Delta(1/R_M)^3$, during solar cycle 23. It is evident that there is exact phase coherence during the two year interval 1998 1999 which we know, from Figure 14, is an interval when periodic sunspot emergence was in phase in both hemispheres or was predominantly due to sunspot emergence in the Southern hemisphere. We infer from this that demonstrating phase coherence of solar activity related variables with the tidal effect of Mercury should be more straightforward during intervals when there is phase coherence of flux emergence in both solar hemispheres.

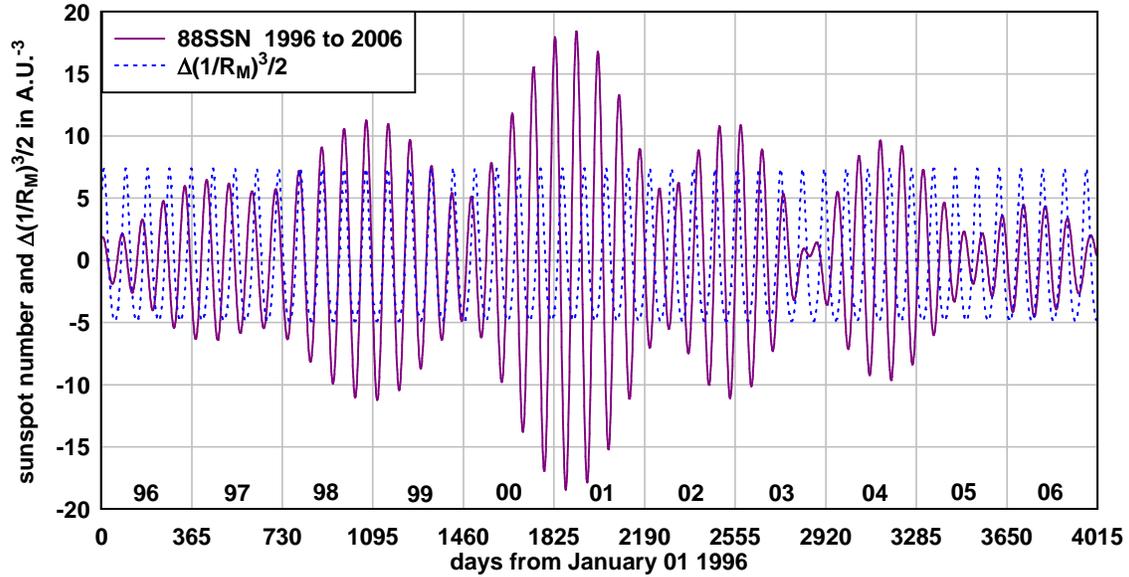

**Figure 15.** The ~ 88 day component of sunspot number (88SSN) for 1996 to 2006 in solar cycle 23 (full line) compared with the variation of the tidal effect of Mercury.

**7.2 Solar wind velocity and cosmic ray flux.**

We expect that other space variables such as solar wind speed and cosmic ray flux might also respond coherently to the tidal effect during intervals when the sunspot emergence in both hemispheres is coherent with the tidal effect of Mercury. Figure 16 compares the ~88 day period components of solar wind velocity and Climax cosmic ray flux with the tidal effect of Mercury for the years 1998, 1999 and 2000. In the middle of this interval, i.e. around 1999, the ~88 day component of solar wind velocity varies in phase with the tidal effect and the ~88 day component of cosmic ray flux varies in anti-phase with the tidal effect. Thus, during this interval there is a relatively straightforward explanation of the observed variations: (1) Magnetic flux emerges on the surface of the Sun in phase with the 88 day period tidal effect. (2) The magnetic flux enters the heliosphere as fast solar wind the speed of which varies in phase with the tidal effect. (3) The fast solar wind reduces the influx of galactic cosmic rays to the inner heliosphere and, as a result, the cosmic ray flux at Earth varies in anti-phase with the Mercury tidal effect. We show below that the ~88 day components of the solar wind velocity and cosmic ray flux amount to a small percentage of the overall level of these components. Thus exact phase coherence as is observed in the case for sunspot area is unlikely.

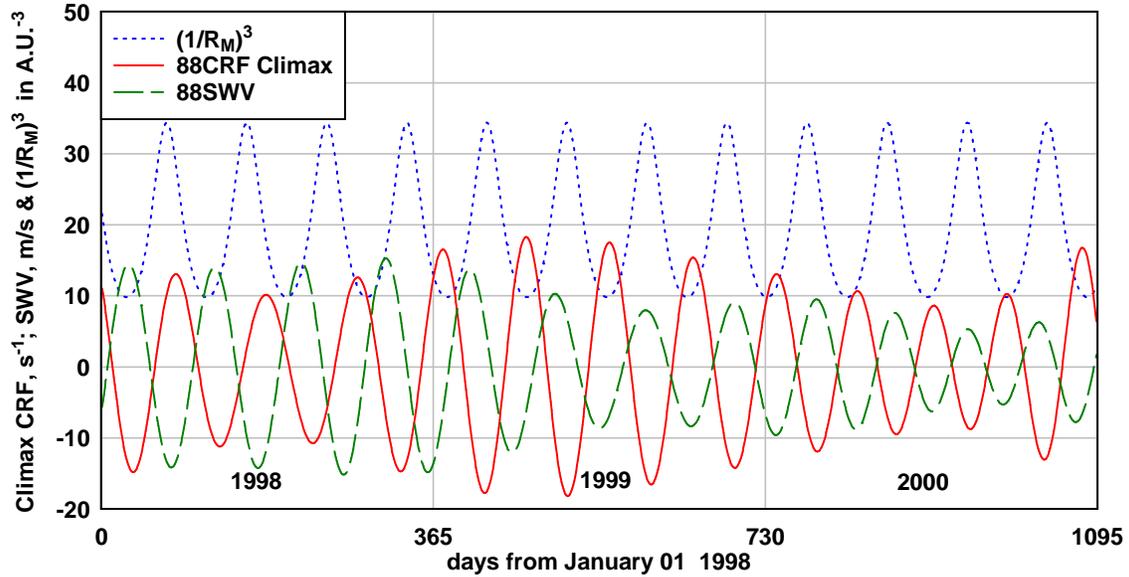

**Figure 16.** The ~88 day period components of solar wind velocity (broken green line) and Climax cosmic ray flux (full red line) compared with the tidal effect of Mercury (blue dots) for 1998 1999 and 2000.

During 1999 the peak to peak variation of the ~ 88 day component of solar wind velocity was, from Figure 16, about 20 m/s during one orbit of Mercury. The average level of solar wind velocity during 1999 is 438 m/s so the percentage variation of solar wind velocity during one orbit of Mercury is 4.5%. Also from Figure 16 the peak to peak variation of the ~88 day period component of cosmic ray flux is 34 s$^{-1}$ during one orbit of Mercury. The average level of Climax cosmic ray flux during 1999 is 4010 s$^{-1}$ so the percentage variation of cosmic ray flux during one orbit of Mercury is 0.8%. For completeness, the percentage variation of the ~88 day component of SSN during one orbit of Mercury in 1999 is 18% and the percentage variation of the ~88 day component of sunspot area South is 30%. Thus we note that, while the percentage contribution of the 88 day period tidal effect to the level of sunspot area and sunspot number is relatively large, 20% - 30%, the percentage contribution of the tidal effect to the level of solar wind speed and cosmic ray flux is relatively small, 1% - 5%. This may be due to the fact that the variation of solar wind speed and cosmic ray flux largely derives from the strength and tilt of the polar magnetic field rather than the effect of the magnetic field associated with the toroidal field and sunspots.

### 7.3 Solar X-ray flares.

Flares erupt from sunspots and the energy output in the form of radiation and charged particles can have a significant impact on communications and electrical infrastructure on Earth. Hence the interest in the possibility that flares may be periodic and that flares may be predictable, Reiger et al (1984), Hung (2007). Interest in "quasi-periodicities" in solar activity began with and continues with studies of quasi-periodicity in flare number. Here we use, as an example of the influence of Mercury on flares, a report by Ballester et al (1999) that compared sunspot area, sunspot number and high energy X-ray flare number during the interval 1979 to 1984 in solar cycle 21, (Ballester et al., 1999, Figure 3). They observed eleven regularly spaced pulses of flare activity during this interval which they interpreted as due to a ~160 day Reiger type "quasi-periodicity" in flare activity. We re-examine the flare activity during 1979 - 1984 from the viewpoint of the findings in this paper. The flare activity reported by Ballester et al (1999) was the combined flare count from both hemispheres therefore we assess the variation of sunspot area in both hemispheres during 1979 – 1984. Figure 17A and 17B show, respectively, the ~176 day period components of sunspot area North and sunspot area South along with the Mercury tidal effect. In the Northern hemisphere, Figure 17A, a strong activation in 1979 – 1983 overlaps a weak activation in 1983 – 1984. The emergence of sunspot area is in anti-phase to every second minimum of the tidal effect during the strong activation. In the Southern hemisphere, Figure 17B, there are two moderately strong activations. The variation in the activation of 1979 – 1981 is in phase with every second peak of the tidal effect and the variation in the second activation is moving to anti-phase with every second minimum of the tidal effect. When the ~176 day period components of sunspot area on the North and South hemispheres are added the result in Figure 17C is obtained. We note that during 1979 – 1981 the sunspot area variations in the North and South differ by one 88 day interval of tidal effect variation and therefore by $\pi/2$ of the 176 day sub harmonic variation; i.e. the variations in the two hemispheres are near in phase and add constructively. In fact examination of Figures 17A and 17B show the variations add constructively throughout the entire interval. The peaks in the combined sunspot area of Figure 17C correspond closely in time to the peaks in high energy X-ray flare number in Figure 3(b) Ballester et al (1999). We infer from this, as did Ballester et al (1999), that flare activity peaked when sunspot area peaked during this interval. However, we note that the combined ~176 day period components of sunspot area in Figure 17C do not show a clear, symmetrical, phase coherence with the tidal effect whereas the separate North and South hemisphere components of sunspot area do. We suggest that this is a further example of the situation where the combined effects of North and South hemisphere activity are difficult to interpret whereas the effects of Northern hemisphere activity and Southern hemisphere activity viewed separately have a straightforward interpretation in terms of the tidal effect.

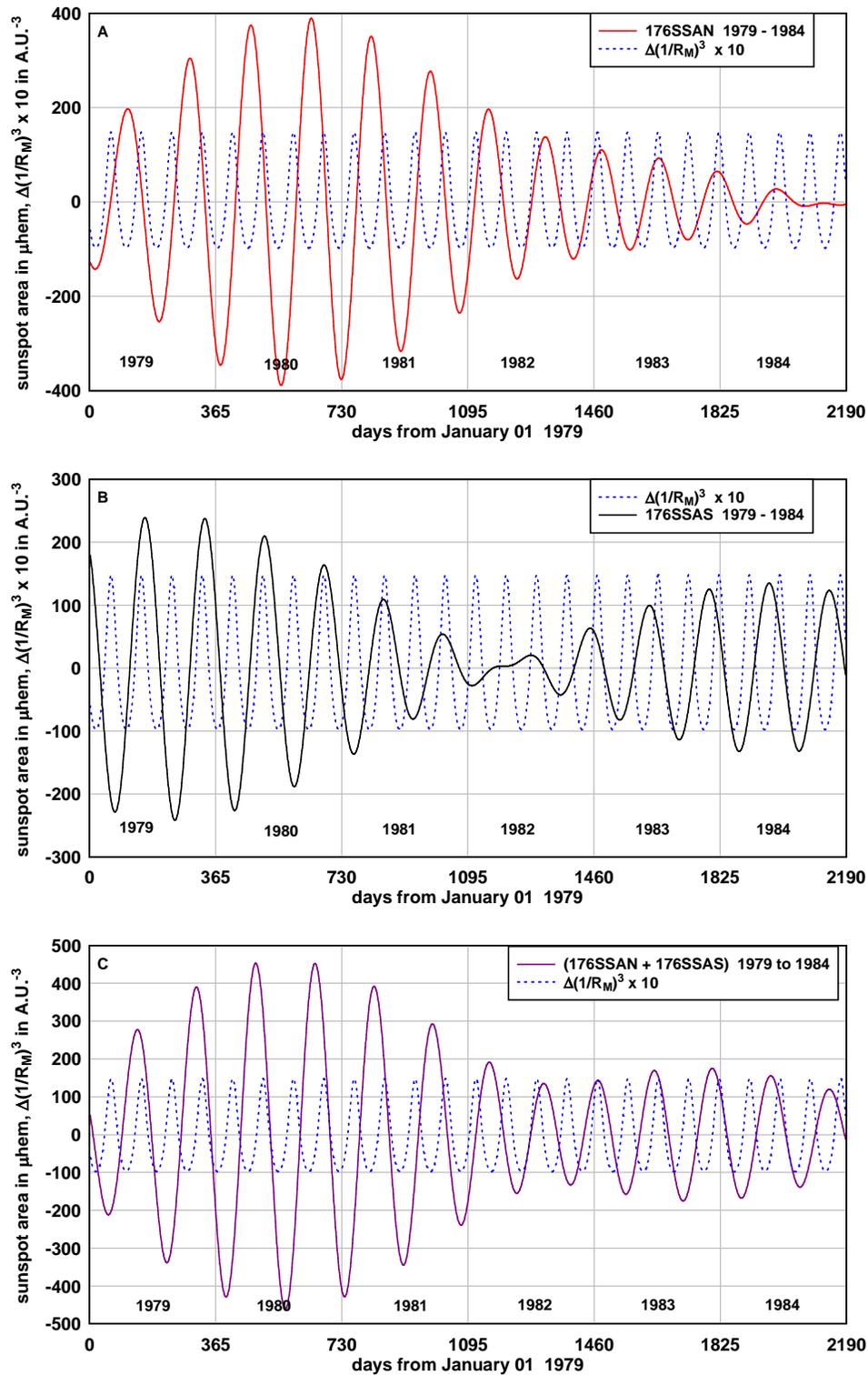

**Figure 17.** (A) Compares the ~ 176 day period variation of sunspot area North with the Mercury tidal effect during 1979 – 1984. (B) Compares the ~ 176 day period variation of sunspot area South with the Mercury tidal effect during 1979 – 1984. (C) Compares the combined variation of the ~176 day components of sunspot area North and sunspot area South with the Mercury tidal effect during 1979 – 1984.

As mentioned above interest in periodicity in flare activity is motivated by the possibility of predicting flares. The results in Figure 17 above and in Ballester et al (1999) Figure 3 indicate that peak flare activity coincides with peaks in sunspot area. It is evident from the results in this paper that peaks in sunspot area are related to the tidal effect of Mercury, the phase of which is known precisely into the future. However, we have shown that both the ~88 day period sunspot area variation and the ~176 day sunspot area variation can exist in two different phase states relative to the tidal effect during discrete activations and exist in a variable phase state relative to the tidal effect during overlapping activations. Thus the possibility of predicting flares based on coherence with the tidal effect seems limited except when long, discrete, in-phase activations can be established to be operating in both hemispheres.

**8. Conclusion**

For many years Reiger type "quasi-periodicities" in solar activity related variables have been studied by spectral analysis methods. However, previous studies did not attempt to establish phase coherence between the periodic variation of solar activity related variables and any driving mechanism as the latter was unknown. In this paper the phase of the proposed driving mechanism, the tidal effect of Mercury, is known precisely and it has been possible to establish phase coherence between the tidal effect and the periodic variation of solar activity related variables over long intervals whenever the condition of discrete activation occurred.

We conclude, therefore, that the observations presented above support a connection between the tidal effect of Mercury, sunspot activity on the Sun and variation of solar flares, solar wind speed and cosmic ray flux. The mechanism that emerges is as follows:

(1) There is a minute tidal elongation of the equatorial surface of the Sun due to the 88 day variation of the orbital radius of Mercury.

(2) Unstable equatorial Rossby wave modes on the Sun that have allowed periods very close to the 88 day period of Mercury and the 176 day first sub harmonic period of Mercury are excited by the tidal variation and grow to amplitudes that are sufficiently large to produce observable periodic variations in the Suns diameter.

(3) The Rossby waves trigger periodic emergence of magnetic flux in the form of sunspots in episodes or activations that last for several years.

(4) An activation lasts until the buoyant magnetic flux in the region of the Sun influenced by a particular Rossby wave mode is depleted.

(5) There may be between two and six activations of sunspot area on a hemisphere of the Sun during a solar cycle with activations on Northern and Southern hemispheres usually different in timing, phase and amplitude.

(6) When activation on a hemisphere is discrete in time from other activations on that hemisphere the periodic variation of sunspot area during activation is phase coherent with the tidal effect of Mercury.

(7) Two states of hemispherical phase coherence are evident. The ~88 day period variation of sunspot area may be in phase or in anti-phase with the tidal effect. The ~176 day period variation of sunspot area may be in phase with every second tidal effect peak or in anti-phase with every second tidal effect minimum.

As a result of this mechanism the combination of the ~88 day period and ~176 day period variations in hemispherical sunspot area account, on average over cycle 23, for about 30% of the short term variation of hemispherical sunspot area i.e. the variation at periods shorter than the ~ 11 year cycle variation and, when the ~88 day period variations of sunspot area in the North and South hemispheres are both phase coherent with the tidal effect of Mercury, phase coherent variations are evident at the few percent level in the solar wind speed and the cosmic ray flux.

A second conclusion is that many of the "quasi-periodicities" previously discovered in the spectrograms of variables connected with sunspot activity are at periods corresponding to sidebands associated with amplitude modulation of sunspot emergence due to the 88 day period tidal acceleration due to Mercury and sub harmonics. The amplitude modulation arises from activations of sunspot emergence lasting several years. Knowing the interval between one activation and the next and the probable $\pi$ phase change between activations allows one to predict fairly accurately the sideband periods, $T_S$, using the simple relation $T_S = 1/(f_1 -/+ f_m)$. As outlined in Section 5 spectrograms of sunspot related variables should often show "quasi-periodicities" at around 85, 95, 160 and 195 days in spectrograms obtained over intervals that encompass two or more activations. As a $\pi$ phase shift in variation between activations results in a minimum in spectrograms at the period of the tidal effect a third conclusion is that it may be more useful to search for deep minima rather than high peaks in the spectrograms of variables connected with solar activity.

A fourth conclusion relates to prediction of sunspot and flare activity. It is evident that once a discrete activation has begun in one hemisphere the strength of subsequent sunspot activity and therefore flare activity in that hemisphere is predictable to some extent. If, for example, it was establishd that a long, discrete activation as in Figure 4 had begun in the northern hemisphere it is reasonable to expect higher flare activity on succeeding intervals of 176 days during the next five years or so. However, longer term prediction is difficult despite the precision of the timing of the tidal effect. This is because the phase of the variation in succeeding activations may be difficult to predict, the duration of activation is quite variable and the phase of the variations in the North and South hemispheres often differ.

A fifth conclusion is that much of the previous work that focused on discovering and interpreting Reiger type "quasi-periodicities" can be usefully reinterpreted with the "quasi-periodicities" viewed either as periodicities associated with the 88 day period of Mercury and the 176, 264, 352, 440 day period sub harmonics or as periodicities corresponding to sidebands of these periods; the periods of the sidebands dependent on the type and duration of activation in operation.

A sixth conclusion is that coherence with planetary tides is relatively easy to demonstrate for hemispherical variables like sunspot area North or sunspot area South, but coherence with global variables like sunspot number or solar wind is difficult to demonstrate because Sun weather in the North and South hemispheres is quite different. The analogy is in trying to understand Earth weather by following global averages. However, the paper indicates how to find the occasional times when the Sun weather in both hemispheres is varying coherently with tidal effects and when global variables should, therefore, also vary coherently with tidal effects.

Some unexplained aspects arising from the present work are: (a), how the minute periodic variations in surface height on the Sun due to the tidal effect of Mercury can excite the growth of Rossby waves to observable amplitude; (b), how Rossby waves trigger the emergence of magnetic flux on the Sun; (c), why emergence occurs over varying intervals of a few years; (d), why, when intervals of emergence or activations are discrete, there is a $\pi$ phase change in the periodic variation of sunspots from one activation to the next; and, (e), given that Mercury influences sunspot emergence, how do the other planets and conjunctions of planets influence sunspot emergence?

**References.**